\documentclass[aps,twocolumn,showpacs,showkeys]{revtex4}

\usepackage{graphics}
\usepackage{bm}
\usepackage{amssymb}

\begin{document}

\title{Infinitesimal propagation equation for decoherence of an OAM entangled biphoton in atmospheric turbulence}
\author{Filippus S. Roux}
\affiliation{CSIR National Laser Centre, P.O. Box 395, Pretoria 0001, South Africa}
\email{fsroux@csir.co.za}

\begin{abstract}
We derive a first order differential equation for the decoherence of an orbital angular momentum entangled biphoton state propagating through a turbulent atmosphere. The derivation is based on the distortion that orbital angular momentum states experience due to propagation through a thin sheet of turbulent atmosphere. This distortion is treated as an infinitesimal transformation leading to a first order differential equation, which we call an infinitesimal propagation equation. The equation is applied to a simple qubit case to show how the entanglement decays.
\end{abstract}


\maketitle

\section{Introduction}

The orbital angular momentum (OAM) states of a photon is an attractive basis for quantum information processing and communication, because it allows a higher dimensional representation of quantum information \cite{zeil1,qkdn,torres}. One application where such a higher dimensional representation can have benefits in terms of information capacity is in free-space quantum communication. Unfortunately, the turbulence in the atmosphere causes the decoherence of entanglement of OAM states. The refractive index fluctuations in the turbulent atmosphere produce random phase modulations that distort the OAM states. This process is similar to the distortion suffered by OAM modes in classical optical beams due to scintillation in turbulence \cite{turboam0,turboam1,turboam2,turboam3,turboam4}.

In this paper we derive a theoretical framework with which one can investigate the decoherence of OAM entanglement in atmospheric turbulence. Some previous investigations of this process \cite{qturb1,qturb3,qturb4} are based on the work of Patterson \cite{turboam1}, who assumed that one can model the turbulent medium with a single phase screen and parameterized the turbulence with one parameter, the Fried parameter \cite{scintbook}. As a result the effects of the propagation, such as beam spreading, intensity scintillation and multiple scattering among different modes, are not incorporated into the model. Neither is this framework able to consider the individual effects of the different scale parameters that are combined into the Fried parameter.

Another approach to investigating the decoherence of OAM entanglement in atmospheric turbulence is to model the turbulent medium as an absorbing and scattering medium \cite{qturb2}. However, this analysis assumes that the turbulence is weak.

The approach that is used in this paper is based on the incremental effect of the turbulent atmosphere on the quantum state (density operator), expressed as a first order differential equation, which can then be solved to obtain the evolution over the entire propagation distance. This approach is reminiscent of a master equation approach, however, here, the derivative is taken with respect to propagation distance instead of time. In this framework one can incorporate any model for the turbulence --- i.e.\ any power spectral density such as Kolmogorov, Tartaskii or von Karman \cite{scintbook}. By implication one can investigate the effect of any of the scale parameters associated with the whole process and not only the Fried parameter. One can apply this framework to cases with arbitrary strong turbulence or strong scintillation.

For the purpose of the derivation, it is assumed that the light is monochromatic, that the beam propagates paraxially and that it has a uniform polarization to allow scalar propagation. The final result can be generalized to polychromatic vector fields. The paraxial approximation would always be valid in practical applications of this framework.

The paper is organized as follows. First we discuss how to treat a photon field that propagates through a linear spatially inhomogeneous medium in Section~\ref{fqua}. Then we discuss the infinitesimal transformation of the momentum space wave functions of such photon fields in Section~\ref{eom}. In Section~\ref{denop} we show how we define the density operator in terms of the OAM and momentum bases. The derivation of the infinitesimal propagation equation (IPE) is shown in Section~\ref{deripe}, with the aid of the ensemble averaging discussed in Appendix~\ref{ensemb}. Then in Section~\ref{solvint} we show how one can treat the different integrals that appear in the IPE, using the generating function of OAM model discussion in Appendix~\ref{oamgen}. In Section~\ref{qubit} a simple example is considered. We discuss the results in Section~\ref{disc} and end with a summary in Section~\ref{summ}.

\section{Field quantization}
\label{fqua}

A dynamical system is generally understood to be one evolving in time. For a quantum system this evolution is unitary, being described by a unitary operator that is given by an exponential operator with the time integrated Hamiltonian as its argument. Within this context the process of decoherence of OAM entanglement in a turbulent atmosphere is not described by a dynamical system. Although the turbulent medium is inhomogeneous it is still assumed to be linear \footnote{One could incorporate nonlinear interactions between the air and the light traversing it to obtain a more exact model for the physics of the problem, but we argue that such an elaborate model is not necessary to capture the predominant mechanism of the decoherence in this process.}, which implies that one can consider each temporal frequency separately, leading to a monochromatic assumption. The temporal behavior of the field is therefore simply given by $\exp(i\omega t)$. A straight forward Hamiltonian approach for this problem is therefore inappropriate \footnote{This is not to say that one cannot use some artificial Hamiltonian that somehow converts the spatial evolution into a temporal evolution. However, this is not our approach here.}. Instead of having a three-dimensional field that evolves in time, our approach in this paper is to ignore the temporal behavior and instead consider a two-dimensional field that changes as it moves along the third spatial dimension.

Due to the monochromatic assumption, which fixes the wavelength of the light $\lambda$, the $z$-component of the propagation vector can be expressed as a function of the other two components
\begin{equation}
k_z(k_x,k_y) = \left( {4\pi^2\over \lambda^2}-k_x^2-k_y^2\right)^{1/2} .
\end{equation}
As a result the momentum basis under the monochromatic approximation becomes a two-dimensional basis $|{\bf K}\rangle$, where the propagation vector ${\bf K} = k_x \hat{x} + k_y \hat{y}$ represents the two-dimensional projection of the three-dimensional propagation vector. It is necessary to include the evanescent momentum states (those for which $k_z$ is imaginary) to ensure that the basis is complete. However, under the paraxial approximation the evanescent part of the field is negligible. We assume that the $z$-direction is the general direction of propagation. Therefore, the two-dimensional momentum basis describes two-dimensional fields on the transverse plane, perpendicular to the propagation direction. In the absence of turbulence one can recover the full three-dimensional field by using Fresnel diffraction theory to propagate the two-dimensional field along $z$. In the presence of turbulence the two-dimensional field transforms in a more complicated way from one plane to another. We formulate the decoherence process in terms of this transformation.

An arbitrary pure state of a monochromatic, uniformly polarized single photon can now be expressed on the transverse plane in terms of the two-dimensional momentum basis
\begin{equation}
|\psi\rangle = \int G({\bf K}) |{\bf K}\rangle\ {{\rm d}^2 K\over 4\pi^2} ,
\end{equation}
where the coefficient function $G({\bf K})$ is the momentum space wave function $\langle {\bf K} | \psi \rangle$. The inverse two-dimensional Fourier transform of the momentum space wave function gives the position space wave function $g(x,y)$ on the transverse plane at $z=0$.

Although the OAM basis (which is equivalent to the two-dimensional momentum basis) is the preferred basis for an analysis where the photons are entangled in term of OAM, the turbulence is modeled in terms of a power spectral density that is defined in terms of a momentum basis. As a result one is forced to have a mixture of both bases in this analysis. We therefore start with the momentum basis and eventually express the OAM basis in terms of it.

\section{Equation of motion and infinitesimal transformation}
\label{eom}

For an interaction-free system the quantum wave function obeys the same equations of motion as the classical field. Therefore, the equation of motion for the position space wave function can be derived from Maxwell's equations. In a source-free region this gives the Helmholtz equation
\begin{equation}
\nabla^2 E({\bf x}) + n^2 k^2 E({\bf x}) = 0 ,
\end{equation}
where $E({\bf x})$ is the scalar part of the electric field (assuming the polarization is uniform and can be ignored), $k$ is the wave number, $n$ is the refractive index and ${\bf x} = x \hat{x} + y \hat{y} + z \hat{z}$. The inhomogeneous medium is represented by a spatially varying index of refraction
\begin{equation}
n = 1 + \delta n ({\bf x}) .
\end{equation}
This variation is very small ($\delta n \ll 1$), which implies that one can approximate the Helmholtz equation as
\begin{equation}
\nabla^2 E({\bf x}) + k^2 E({\bf x}) + 2 \delta n({\bf x}) k^2 E({\bf x}) = 0 .
\end{equation}
Furthermore, we assume that the beam is paraxial and propagates in the $z$-direction. So we define
\begin{equation}
E({\bf x}) = g({\bf x}) \exp(-i k z) ,
\end{equation}
which then leads to the paraxial wave equation with the extra inhomogeneous medium term
\begin{equation}
\nabla_T^2 g({\bf x}) - i 2 k \partial_z g({\bf x}) + 2 \delta n({\bf x}) k^2 g({\bf x}) = 0 ,
\end{equation}
where $\nabla_T$ is the transverse part of the gradient operator.

Now we replace $g({\bf x})$ with its two-dimensional inverse Fourier transform,
\begin{equation}
g({\bf x}) = \int G({\bf K},z) \exp(-i {\bf K}\cdot {\bf x})\ {{\rm d}^2 K\over 4\pi^2} ,
\end{equation}
which contains the angular spectrum of the optical field $G({\bf K},z)$. The latter also represents the momentum space wave function for the purpose of the quantum analysis. Then we obtain
\begin{equation}
\partial_z G({\bf K},z) = {i\over 2k} |{\bf K}|^2 G({\bf K},z) - i k N({\bf K},z) \star G({\bf K},z) ,
\label{spekprop}
\end{equation}
where $\star$ indicates convolution and $N({\bf K},z)$ is the two-dimensional Fourier transform of $\delta n({\bf x})$. This equation represents the infinitesimal transformation of the momentum space wave function during propagation. It forms the basis of the derivation of the IPE.

\section{Density operator in the OAM basis}
\label{denop}

For the purpose of this derivation we first consider a single photon and then generalize the result for the case of two photons. The density operator of an arbitrary single photon state can be expressed in the OAM base by
\begin{equation}
\rho = \sum_{m,n} |m\rangle\ \rho_{m,n}\  \langle n| ,
\label{rhooam}
\end{equation}
where we use a single index to denote both the indices of the OAM states, $m\equiv \{l,r\}$. (See Appendix \ref{oamgen} for a discussion of the OAM modes.) Unless stated otherwise, each OAM index used in the subsequent derivation always represents both the indices associated with a particular OAM mode. Each of these OAM states can be expanded in terms of the two-dimensional momentum basis
\begin{equation}
|m\rangle = \int G_m({\bf K}) |{\bf K}\rangle\ {{\rm d}^2 K\over 4\pi^2} ,
\end{equation}
leading to the following expression for the density operator in terms of the two-dimensional momentum basis
\begin{eqnarray}
\rho(z) & = & \sum_{m,n} \int |{\bf K}_1\rangle\  G_m({\bf K}_1,z) \rho_{m,n}  \nonumber \\ & & \times G_n^*({\bf K}_2,z) \langle{\bf K}_2|\ {{\rm d}^2 K_1\over 4\pi^2} {{\rm d}^2 K_2\over 4\pi^2} ,
\label{spk}
\end{eqnarray}
where the dependence on $z$ is shown explicitly to make it apparent that this expression for $\rho$ is only valid on a transverse plane for a specific value of $z$. Here, and also later, we use only one integral sign to represent several $K$-space integrals. By evaluating the summations in Eq.~(\ref{spk}), one obtains the definition for the density operator in terms of the two-dimensional momentum basis,
\begin{equation}
\rho(z) = \int |{\bf K}_1\rangle \rho({\bf K}_1,{\bf K}_2,z) \langle {\bf K}_2|\ {{\rm d}^2 K_1\over 4\pi^2} {{\rm d}^2 K_2\over 4\pi^2} ,
\label{rhok}
\end{equation}
where
\begin{equation}
\rho({\bf K}_1,{\bf K}_2,z) = \sum_{m,n} G_m({\bf K}_1,z) \rho_{m,n} G_n^*({\bf K}_2,z) ,
\label{oamk}
\end{equation}
which implies that
\begin{equation}
\rho_{m,n} = \int G_m^*({\bf K}_1,z) \rho({\bf K}_1,{\bf K}_2,z) G_n({\bf K}_2,z)\ {{\rm d}^2 K_1\over 4\pi^2} {{\rm d}^2 K_2\over 4\pi^2} .
\label{koam}
\end{equation}
Since the two-dimensional momentum basis and the OAM basis are completely equivalent, the definitions in Eqs.~(\ref{rhooam}) and (\ref{rhok}) are also completely equivalent and Eqs.~(\ref{oamk}) and (\ref{koam}) indicate how one can transform from one to the other.

For two photons the density operator in Eq.~(\ref{spk}) can be generalized to become
\begin{eqnarray}
\rho(z) & = & \sum_{m,n,p,q} \int |{\bf K}_1\rangle |{\bf K}_3\rangle G_m({\bf K}_1,z) G_p({\bf K}_3,z) \nonumber \\ & & \times \rho_{m,n,p,q}\ G_n^*({\bf K}_2,z) G_q^*({\bf K}_4,z) \langle{\bf K}_2| \langle{\bf K}_4| \nonumber \\ & & \times {{\rm d}^2 K_1\over 4\pi^2} {{\rm d}^2 K_2\over 4\pi^2} {{\rm d}^2 K_3\over 4\pi^2} {{\rm d}^2 K_4\over 4\pi^2} .
\label{digt4}
\end{eqnarray}

To obtain the full three-dimensional expression for the density operator in free-space (without turbulence) one can use Fresnel diffraction theory to determine the expression at any other value of $z$. In the presence of turbulence the expression for the density operator is only valid on a specific transverse plane, and it needs to be transformed according to Eq.~(\ref{spekprop}) from plane to plane.

Note that inside the expression the $z$-dependence is carried by the momentum space wave functions and not by the density matrix elements. This is because the transformation of the density operator during propagation over an infinitesimal distance through turbulence is caused by the distortion of the momentum space wave functions. After such an infinitesimal propagation these momentum space wave functions no longer represent the Fourier transforms of the original modes. One needs to re-expand these distorted wave functions in terms of the momentum space wave functions of the OAM modes and incorporate the expansion coefficients in the density matrix elements. Thereby one can transfer the $z$-dependence to the density matrix elements.

\begin{figure}[ht]
\centerline{\scalebox{1}{\includegraphics{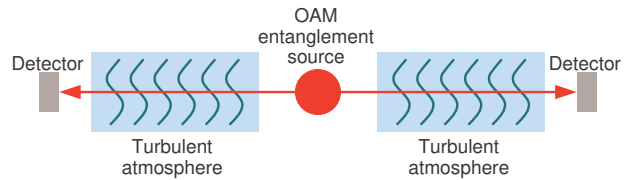}}}
\caption{Diagram showing how the OAM entangled biphoton propagates through turbulent media toward the two detectors.}
\label{bron}
\end{figure}

For notational convenience we represent the product of momentum space wave functions that appear in Eq.~(\ref{digt4}), as a single function,
\begin{eqnarray}
& & G_m({\bf K}_1,z) G_p({\bf K}_3,z) G_n^*({\bf K}_2,z) G_q^*({\bf K}_4,z) \nonumber \\ & = & F_{m,n,p,q}({\bf K}_1,{\bf K}_2,{\bf K}_3,{\bf K}_4,z) .
\end{eqnarray}
The expression for $\rho$ is now given by
\begin{eqnarray}
\rho(z) & = & \sum_{m,n,p,q} \int |{\bf K}_1\rangle |{\bf K}_3\rangle\ \rho_{m,n,p,q} \nonumber \\ & & \times F_{m,n,p,q}({\bf K}_1,{\bf K}_2,{\bf K}_3,{\bf K}_4,z) \nonumber \\ & & \times \langle{\bf K}_2|\langle{\bf K}_4|\ {{\rm d}^2 K_1\over 4\pi^2}  {{\rm d}^2 K_2\over 4\pi^2} {{\rm d}^2 K_3\over 4\pi^2} {{\rm d}^2 K_4\over 4\pi^2} .
\label{digt4s}
\end{eqnarray}
Since $F_{m,n,p,q}$ carries the only $z$-dependence in the expression for the density operator, we focus on how it transforms during infinitesimal propagation. At the end we apply the transformation to the expression for $\rho$.

\section{Derivation of the IPE}
\label{deripe}

We are ready to consider the steps in the derivation of the IPE. For this purpose we consider the scenario where a pair of photons (biphoton), which are entangled in terms of the OAM basis, both propagate through a turbulent atmosphere, as shown in Fig.~\ref{bron}. It is assumed that the turbulences seen by the respective photons are mutually uncorrelated.

\subsection{Infinitesimal transformation}

The transformation that is caused by an infinitesimal propagation is obtained by substituting Eq.~(\ref{spekprop}) into the infinitesimally propagated version of $F_{m,n,p,q}$, which is obtained by setting $z=z+dz$ and then expanding the result to subleading order in $dz$. The resulting expression is turned into a differential equation by taking the limit $dz\rightarrow 0$,
\begin{widetext}
\begin{eqnarray}
\partial_z F_{m,n,p,q}({\bf K}_1,{\bf K}_2,{\bf K}_3,{\bf K}_4,z) & = & {i\over 2k} (|{\bf K}_1|^2-|{\bf K}_2|^2+|{\bf K}_3|^2-|{\bf K}_4|^2) F_{m,n,p,q}({\bf K}_1,{\bf K}_2,{\bf K}_3,{\bf K}_4,z) \nonumber \\
& & - i k \left[
\int N_1({\bf K}_1-{\bf K},z) F_{m,n,p,q}({\bf K},{\bf K}_2,{\bf K}_3,{\bf K}_4,z)\  {{\rm d}^2 K\over 4\pi^2} \right. \nonumber \\
& & - \int N_1^*({\bf K}_2-{\bf K},z) F_{m,n,p,q}({\bf K}_1,{\bf K},{\bf K}_3,{\bf K}_4,z)\  {{\rm d}^2 K\over 4\pi^2} \nonumber \\
& & + \int N_2({\bf K}_3-{\bf K},z) F_{m,n,p,q}({\bf K}_1,{\bf K}_2,{\bf K},{\bf K}_4,z)\  {{\rm d}^2 K\over 4\pi^2} \nonumber \\
& & \left. - \int N_2^*({\bf K}_4-{\bf K},z) F_{m,n,p,q}({\bf K}_1,{\bf K}_2,{\bf K}_3,{\bf K},z)\ {{\rm d}^2 K\over 4\pi^2} \right] .
\label{ensf0}
\end{eqnarray}
\end{widetext}
This differential equation still contains the $N_n$'s, which are random functions. One needs to perform an ensemble averaging to remove the randomness. However, if one would perform such an ensemble averaging on Eq.~(\ref{ensf0}) all the terms containing $N_n$'s would fall away, because $\langle N_n \rangle = 0$. So we proceed as follows. Integrating Eq.~(\ref{ensf0}), one obtains $F_{m,n,p,q}$ in terms of previous versions of itself. Next, one substitutes the resulting integral equation back into itself repeatedly to obtain an infinite series (Dyson expansion). The ensemble average of the result would remove many terms, because
\begin{equation}
\langle N_n \rangle = \langle N_n N_m^{(*)} \rangle = 0 ~~~~~~~ {\rm for} ~~~~ m\neq n .
\end{equation}
The resulting integral equation, up to second order in $N_n$, showing only the terms that survives the ensemble averaging, is given by
\begin{widetext}
\begin{eqnarray}
F_{m,n,p,q}({\bf K}_1,{\bf K}_2,{\bf K}_3,{\bf K}_4,z) & = & F_{m,n,p,q}({\bf K}_1,{\bf K}_2,{\bf K}_3,{\bf K}_4,z_0) \nonumber \\
& & + (z-z_0) {i\over 2k} (|{\bf K}_1|^2-|{\bf K}_2|^2+|{\bf K}_3|^2-|{\bf K}_4|^2) F_{m,n,p,q}({\bf K}_1,{\bf K}_2,{\bf K}_3,{\bf K}_4,z_0)  \nonumber \\
& & - k^2\int_{z_0}^{z} \int_{z_0}^{z_1} \int \langle N_1({\bf K}_1-{\bf K},z_1) N_1({\bf K}-{\bf K}_0,z_2) \rangle F_{m,n,p,q}({\bf K}_0,{\bf K}_2,{\bf K}_3,{\bf K}_4,z_0) \nonumber \\
& & - \langle N_1({\bf K}_1-{\bf K},z_1) N_1^*({\bf K}_2-{\bf K}_0,z_2) \rangle F_{m,n,p,q}({\bf K},{\bf K}_0,{\bf K}_3,{\bf K}_4,z_0) \nonumber \\
& & - \langle N_1^*({\bf K}_2-{\bf K},z_1) N_1({\bf K}_1-{\bf K}_0,z_2) \rangle F_{m,n,p,q}({\bf K}_0,{\bf K},{\bf K}_3,{\bf K}_4,z_0) \nonumber \\
& & + \langle N_1^*({\bf K}_2-{\bf K},z_1) N_1^*({\bf K}-{\bf K}_0,z_2) \rangle F_{m,n,p,q}({\bf K}_1,{\bf K}_0,{\bf K}_3,{\bf K}_4,z_0) \nonumber \\
& & + \langle N_2({\bf K}_3-{\bf K},z_1) N_2({\bf K}-{\bf K}_0,z_2) \rangle F_{m,n,p,q}({\bf K}_1,{\bf K}_2,{\bf K}_0,{\bf K}_4,z_0) \nonumber \\
& & - \langle N_2({\bf K}_3-{\bf K},z_1) N_2^*({\bf K}_4-{\bf K}_0,z_2) \rangle F_{m,n,p,q}({\bf K}_1,{\bf K}_2,{\bf K},{\bf K}_0,z_0) \nonumber \\
& & - \langle N_2^*({\bf K}_4-{\bf K},z_1) N_2({\bf K}_3-{\bf K}_0,z_2) \rangle F_{m,n,p,q}({\bf K}_1,{\bf K}_2,{\bf K}_0,{\bf K},z_0) \nonumber \\
& & + \langle N_2^*({\bf K}_4-{\bf K},z_1) N_2^*({\bf K}-{\bf K}_0,z_2) \rangle F_{m,n,p,q}({\bf K}_1,{\bf K}_2,{\bf K}_3,{\bf K}_0,z_0)\ {{\rm d}^2 K_0\over 4\pi^2}\ {{\rm d}^2 K\over 4\pi^2}\ {\rm d} z_2\ {\rm d} z_1 \nonumber \\
\label{ensf}
\end{eqnarray}
Note that the only $z_1$- and $z_2$-dependences appear in the $N_n$'s. In Appendix \ref{ensemb} it is shown that
\begin{equation}
\int_{z_0}^{z} \int_{z_0}^{z_1} \langle N_n({\bf K}_1,z_2) N_n^*({\bf K}_2,z_1) \rangle\ {\rm d} z_2\ {\rm d} z_1 = 2\pi^2 dz \delta({\bf K}_1-{\bf K}_2) \Phi_1({\bf K}_1) ,
\label{verw3}
\end{equation}
where we set $z-z_0=dz$. We now use Eq.~(\ref{verw3}) to simplify Eq.~(\ref{ensf})  and then take the limit $dz\rightarrow 0$, to turn it into a differential equation again,
\begin{eqnarray}
\partial_z F_{m,n,p,q}({\bf K}_1,{\bf K}_2,{\bf K}_3,{\bf K}_4,z)
& = & {i \over 2k} (|{\bf K}_1|^2-|{\bf K}_2|^2+|{\bf K}_3|^2-|{\bf K}_4|^2) F_{m,n,p,q}({\bf K}_1,{\bf K}_2,{\bf K}_3,{\bf K}_4,z_0) \nonumber \\
& & - 2 k^2 F_{m,n,p,q}({\bf K}_1,{\bf K}_2,{\bf K}_3,{\bf K}_4,z_0) \int \Phi_1({\bf K})\ {{\rm d}^2 K\over 4\pi^2} \nonumber \\
& & + k^2 \int \Phi_1({\bf K}) F_{m,n,p,q}({\bf K}_1-{\bf K},{\bf K}_2-{\bf K},{\bf K}_3,{\bf K}_4,z_0)\
 {{\rm d}^2 K\over 4\pi^2} \nonumber \\
& & + k^2 \int \Phi_1({\bf K}) F_{m,n,p,q}({\bf K}_1,{\bf K}_2,{\bf K}_3-{\bf K},{\bf K}_4-{\bf K},z_0)\
 {{\rm d}^2 K\over 4\pi^2}  .
\label{ensf1}
\end{eqnarray}
If one substitutes Eq.~(\ref{ensf1}) into the $z$-derivative of Eq.~(\ref{digt4s}) one would obtain a first order differential equation for the density operator. However, we are interested in the transformation of the individual density matrix elements.

\subsection{Extraction of matrix elements}

To express the transformation of the density operator due to the infinitesimal propagation through a turbulent atmosphere in terms of the density matrix elements, one needs to extract the matrix elements from the density operator, using the trace
\begin{equation}
\partial_z \rho_{u,v,r,s}(z) = {\rm trace} \left\{ \partial_z \rho(z) |v\rangle |s\rangle \langle u| \langle r| \right\} ,
\label{trace0}
\end{equation}
where the operator that selects a particular matrix element in the OAM basis is given by
\begin{equation}
|v\rangle |s\rangle \langle u| \langle r| = \int |{\bf K}_8\rangle |{\bf K}_6\rangle\ F_{u,v,r,s}^*({\bf K}_5,{\bf K}_6,{\bf K}_7,{\bf K}_8,z)
\langle{\bf K}_7|\langle{\bf K}_5|\ {{\rm d}^2 K_5\over 4\pi^2}  {{\rm d}^2 K_6\over 4\pi^2} {{\rm d}^2 K_7\over 4\pi^2} {{\rm d}^2 K_8\over 4\pi^2} .
\label{digt5}
\end{equation}
Substituting Eqs.~(\ref{digt4s}) and (\ref{digt5}) into Eq.~(\ref{trace0}) we obtain
\begin{eqnarray}
\partial_z \rho_{u,v,r,s}(z) & = &  {\rm trace} \left\{ \sum_{m,n,p,q} \int |{\bf K}_1\rangle |{\bf K}_3\rangle\ \rho_{m,n,p,q} \partial_z F_{m,n,p,q}({\bf K}_1,{\bf K}_2,{\bf K}_3,{\bf K}_4,z) \langle{\bf K}_2|\langle{\bf K}_4|\ {{\rm d}^2 K_1\over 4\pi^2} \right.
{{\rm d}^2 K_2\over 4\pi^2} {{\rm d}^2 K_3\over 4\pi^2} {{\rm d}^2 K_4\over 4\pi^2} \nonumber \\
& & \left. \times \int |{\bf K}_8\rangle |{\bf K}_6\rangle\ F_{u,v,r,s}^*({\bf K}_5,{\bf K}_6,{\bf K}_7,{\bf K}_8,z) \langle{\bf K}_7|\langle{\bf K}_5|\ {{\rm d}^2 K_5\over 4\pi^2}  {{\rm d}^2 K_6\over 4\pi^2} {{\rm d}^2 K_7\over 4\pi^2} {{\rm d}^2 K_8\over 4\pi^2} \right \} \nonumber \\
& = & \sum_{m,n,p,q} \rho_{m,n,p,q} \int \partial_z F_{m,n,p,q}({\bf K}_1,{\bf K}_2,{\bf K}_3,{\bf K}_4,z) \nonumber \\
& & \times F_{u,v,r,s}^*({\bf K}_1,{\bf K}_2,{\bf K}_3,{\bf K}_4,z) {{\rm d}^2 K_1\over 4\pi^2} {{\rm d}^2 K_2\over 4\pi^2} {{\rm d}^2 K_3\over 4\pi^2} {{\rm d}^2 K_4\over 4\pi^2}  ,
\label{digt6}
\end{eqnarray}
where the last expression is obtained because of the orthogonality of the momentum basis
\begin{equation}
\langle {\bf K}_1 | {\bf K}_2 \rangle = 4\pi^2 \delta_2 \left( {\bf K}_1 - {\bf K}_2 \right) ,
\label{ortomom}
\end{equation}
as applied for the separate photons. Substituting Eq.~(\ref{ensf1}) into Eq.~(\ref{digt6}), we obtain
\begin{eqnarray}
\partial_z \rho_{u,v,r,s}(z) & = &  \sum_{m,n,p,q} \rho_{m,n,p,q} \int \left[ {i \over 2k} (|{\bf K}_1|^2-|{\bf K}_2|^2+|{\bf K}_3|^2-|{\bf K}_4|^2) F_{m,n,p,q}({\bf K}_1,{\bf K}_2,{\bf K}_3,{\bf K}_4,z_0) \right. \nonumber \\
& & - 2 k^2 F_{m,n,p,q}({\bf K}_1,{\bf K}_2,{\bf K}_3,{\bf K}_4,z_0) \int \Phi_1({\bf K})\ {{\rm d}^2 K\over 4\pi^2} \nonumber \\
& & + k^2 \int \Phi_1({\bf K}) F_{m,n,p,q}({\bf K}_1-{\bf K},{\bf K}_2-{\bf K},{\bf K}_3,{\bf K}_4,z_0)\
 {{\rm d}^2 K\over 4\pi^2} \nonumber \\
& & \left. + k^2 \int \Phi_1({\bf K}) F_{m,n,p,q}({\bf K}_1,{\bf K}_2,{\bf K}_3-{\bf K},{\bf K}_4-{\bf K},z_0)\
 {{\rm d}^2 K\over 4\pi^2} \right] \nonumber \\ & &  \times F_{u,v,r,s}^*({\bf K}_1,{\bf K}_2,{\bf K}_3,{\bf K}_4,z) \ {{\rm d}^2 K_1\over 4\pi^2}  {{\rm d}^2 K_2\over 4\pi^2} {{\rm d}^2 K_3\over 4\pi^2} {{\rm d}^2 K_4\over 4\pi^2} .
\label{ensf2}
\end{eqnarray}
\end{widetext}

\subsection{Final expression}

Due to the orthogonality of the momentum space wave functions of the OAM basis,
\begin{equation}
\int G_m({\bf K},z) G_n^*({\bf K},z)\ {{\rm d}^2 K\over 4\pi^2} = \delta_{m,n} ,
\label{ortowf}
\end{equation}
one can simplify Eq.~(\ref{ensf2}). The resulting first order differential equation, which represents the IPE, is given by
\begin{eqnarray}
\partial_z \rho_{u,v,r,s}(z) & = & S_{m,u}(z) \rho_{m,v,r,s} - S_{v,n}(z) \rho_{u,n,r,s}
\nonumber \\ & & + S_{p,r}(z) \rho_{u,v,p,s} - S_{s,q}(z) \rho_{u,v,r,q}
\nonumber \\ & & + L_{m,n,u,v}(z) \rho_{m,n,r,s} + L_{p,q,r,s}(z) \rho_{u,v,p,q}
\nonumber \\ & &  - 2 L_T \rho_{u,v,r,s} ,
\label{ensf3}
\end{eqnarray}
where repeated indices imply summation and the following definitions were made:
\begin{equation}
S_{x,y}(z) = {i \over 2k} \int |{\bf K}|^2 G_x({\bf K},z) G_y^*({\bf K},z)\ {{\rm d}^2 K\over 4\pi^2} ,
\label{defs}
\end{equation}
\begin{equation}
L_T = k^2 \int \Phi_1({\bf K})\ {{\rm d}^2 K\over 4\pi^2} ,
\label{deflt}
\end{equation}
and
\begin{equation}
L_{m,n,u,v}(z) = k^2 \int \Phi_1({\bf K}) W_{m,u}({\bf K},z) W_{n,v}^*({\bf K},z)\ {{\rm d}^2 K\over 4\pi^2} ,
\label{defl}
\end{equation}
with
\begin{equation}
W_{x,y}({\bf K},z) = \int G_x({\bf K}_1,z) G_y^*({\bf K}_1-{\bf K},z)\ {{\rm d}^2 K_1\over 4\pi^2} .
\label{defw}
\end{equation}

The first four terms of Eq.~(\ref{ensf3}) are the non-dissipative terms, representing the free-space propagation process. The last three terms of Eq.~(\ref{ensf3}) are the dissipative terms, representing the scattering of OAM modes into other OAM modes due to the turbulence. One can separate the dissipative terms into separate pairs for the two photons, each having its own $L_{m,n,u,v}$ and $L_T$ terms.

The IPE in Eq.~(\ref{ensf3}), together with Eqs.~(\ref{defs})--(\ref{defw}), is the main result of this paper. In general, Eq.~(\ref{ensf3}) represents an infinite set of coupled first-order differential equations. Even if the initial state contains only a few lower order modes, the turbulence will cause these modes to be coupled into all other modes. Subsequently, the other modes will couple back into the original modes. Truncating the set of equations, one inevitably excludes part of the coupling among all the different modes. However, this coupling should become progressively smaller for higher order modes. Hence, one may be able to truncate the set at some point while retaining the dominant inter-modal coupling.

\section{Solving the integrals}
\label{solvint}

The momentum space generating function, discussed in Appendix~\ref{oamgen}, and given by
\begin{eqnarray}
{\cal F} \{G\} & = & \frac{\pi}{1+w}\exp \left[ { i\pi (a+ib)p \over 1+w } + {i\pi (a-ib)q \over 1+w } \right. \nonumber \\ & & \left. - {\pi^2 (a^2+b^2)\Omega(t,w) \over 1+w } \right]
\label{msgf0}
\end{eqnarray}
is now used to evaluate the integrals for Eqs.~(\ref{defs})-(\ref{defw}).

The formalism allows one to use any power spectral density $\Phi_0({\bf k})$ for the turbulence. Here we neglect the effect of the inner scales, and use the von Karman power spectral density \cite{scintbook},
\begin{equation}
\Phi_0({\bf k}) = \frac{0.033 C_n^2}{(|{\bf k}|^2+\kappa_0^2)^{11/6}} = \Phi_1(K) ,
\label{karman}
\end{equation}
where $C_n^2$ is the refractive index structure constant for the turbulence and $\kappa_0$ is inversely proportional to the outer scale of the turbulence. The outer scale helps to regularize the integrals, but in the limit of large outer scale it disappears from the final expressions. Since the power spectral density only depends on the magnitude of ${\bf k}$, one can set $k_z=0$ as discussed in Appendix~\ref{ensemb}, so that ${\bf k}$ is replaced by $K=|{\bf K}|$.

\subsection{Free-space propagation term}
\label{fsp}

First we consider the integral in Eq.~(\ref{defs}), which represents the free-space propagation. After evaluating the integral and removing the superfluous mixed terms containing a $p$ times a $q$, we obtain a generating function for the $S$-integral,
\begin{eqnarray}
S_G(z) & = & {i\lambda(1+w_m)(1+w_n) \over 8(1-w_m w_n)^3} \exp\left[ {p_m p_n + q_m q_n \over 2 (1-w_m w_n) } \right] \nonumber \\ & & \times \left[ 2 (1-w_m w_n) + p_m p_n + q_m q_n \right] ,
\label{ints}
\end{eqnarray}
where $p_m$, $p_n$, $q_m$, $q_n$, $w_m$, and $w_n$ are the generating function parameters, associated with the $m$- and $n$-indices. Since the $p$'s (or $q$'s) always appear in products for the $m$- and $n$-indices, the result implies an orthogonality condition for the azimuthal indices. The same is not true for the radial indices --- one finds that the result is non-zero when the radial indices differ by either 0 or 1. When the difference is 0 the final result for explicit modes can be expressed as
\begin{equation}
S_{m,n}(z) = {i (1+|l|+2r) \over 2 z_R} ,
\label{ewes}
\end{equation}
where the azimuthal indices are indicated by $l_m=l_n=l$ and the radial indices are indicated by $r_m=r_n=r$. When the difference between the radial indices is equal to 1 the result for explicit modes is given by
\begin{equation}
S_{m,n}(z) = {i (1+|l|+r)^{1/2} (1+r)^{1/2} \over 2 z_R}
\label{onewes}
\end{equation}
where the azimuthal indices are again indicated by $l_m=l_n=l$, but now the radial indices are indicated by $r=(r_m+r_n-1)/2$.

\subsection{Divergent dissipative term}
\label{ddt}

Next we consider the integral for the dissipative term given in Eq.~(\ref{deflt}). Substituting Eq.~(\ref{karman}) into Eq.~(\ref{deflt}), one obtains,
\begin{equation}
L_T = 0.1244 C_n^2 \lambda^{-2} \kappa_0^{-5/3} .
\label{intlt}
\end{equation}
Hence, $L_T$ diverges for large outer scale ($\kappa_0\rightarrow 0$). However, we find that these terms are canceled by similar terms coming from $L_{m,n,u,v}(z)$.

\subsection{Modal correlation functions}
\label{mcf}

The integral in Eq.~(\ref{defw}) represents the correlation between the momentum space wave functions of the OAM modes. To evaluate this integral we express the generating function of Eq.~(\ref{msgf0}) in polar momentum space coordinates, so that $k_x+ik_y=K\exp(i\phi)$. The result of the integration, left as a generating function, is then given by
\begin{eqnarray}
W_G(K,\phi,z) & = & {\pi \over 2(1-w_m w_n)}\exp\left[ {p_m p_n + q_m q_n\over 2(1-w_m w_n)} \right. \nonumber \\
& & + {i K\exp(i\phi)(p_m\zeta_n^*+q_n\zeta_m)\eta\over 2(1-w_m w_n)} \nonumber \\
& & + {i K\exp(-i\phi)(p_n\zeta_m+q_m\zeta_n^*)\eta\over 2(1-w_m w_n)} \nonumber \\
& & \left. - {K^2 \zeta_m \zeta_n^* \eta^2 \over 2(1-w_m w_n)} \right]
\label{intw}
\end{eqnarray}
where $\zeta_x = z_R-iz - w_x(z_R+iz)$ and $\eta=\lambda/\omega_0$. If we evaluate the azimuthal indices explicitly while leaving the radial indices implicit in terms of the parameter $w_m$ and $w_n$, one can express the correlation function as
\begin{eqnarray}
W_{rG}(K,\phi,z) & = & {\exp(-X)\exp[i(m-n)\phi] \overline{E}_n^{|n|} E_m^{|m|}\over (1-w_m w_n)} \nonumber \\
 & & \times \left[ {r_n!\over(|n|+r_n)!}\right]^{1/2} \left[{r_m!\over(|m|+r_m)!}\right]^{1/2} \nonumber \\
 & & \times \sum_{s=0}^{M(m,n)} {|m|!|n|! (-X)^{-s} \over (|m|-s)! (|n|-s)! s!}
\label{somw}
\end{eqnarray}
where $m(=l_m)$ and $n(=l_n)$ are the azimuthal indices; $r_m$ and $r_n$ are their associated radial indices; and
\begin{eqnarray}
M(m,n) & = &  {1\over 2} \left( |m|+|n|-|m-n| \right) \label{parmm} \\
X & = & {K^2 \zeta_m \zeta_n^* \eta^2 \over 2(1-w_m w_n)} \label{parmx} \\
E_m & = & {i K \zeta_m \eta \over \sqrt{2}(1-w_m w_n)} \label{parme} \\
\overline{E}_n & = & {i K \zeta_n^* \eta \over \sqrt{2}(1-w_m w_n)}  \label{parmes}.
\end{eqnarray}
Note that $W_{rG}$ still represents a generating function with respect to the radial indices.

\subsection{General dissipative term}
\label{gdt}

The two-dimensional integration in Eq.~(\ref{defl}) can be separated into a radial and angular integral in momentum space polar coordinates. Since $\Phi_1(K)$ only depends on the radial coordinate ($K=|{\bf K}|$) the integral over $\phi$ only involves the $\phi$-dependencies of the $W$'s, as expressed in Eq.~(\ref{somw}). The combined $\phi$-dependencies of the two $W$'s is given by $\exp[i(m-u-n+v)\phi]$, where $m$, $n$, $u$ and $v$ are the azimuthal indices of all the modes involved. The result of the angular integral is zero unless $m-u-n+v=0$, in which case the result is the product of the two $W$'s without the $\phi$-dependent exponentials, times $2\pi$. As a result many of the elements in $L_{mnuv}$ vanish.

The result of the remaining $K$-integral is too complicated to express as a single closed form expression. However, one can consider the result on a term-by term basis. These terms all have the form
\begin{equation}
f_m(K) = {A \exp(-B K^2) K^{2m}\over (K^2+\kappa_0^2)^{11/6}} ,
\label{form0}
\end{equation}
where $m$ is a non-negative integer (not to be confused with the combined OAM indices used before), $A$ contains all the multiplicative parameters from Eqs.~(\ref{somw}) and (\ref{karman}), and $B$ is a parameter composed of the parameters in the exponent of Eq.~(\ref{somw}), as given in Eq.~(\ref{parmx}).

If $m=0$ the integral over $f_m(K)$ diverges as $\kappa_0 \rightarrow 0$. In the limit of small $\kappa_0$ the leading terms are,
\begin{equation}
\int_{-\infty}^{\infty} f_0(K) K\ {\rm d}K \approx L_T - {\pi^{1/2} \over 6\Gamma(2/3)} {C_n^2 \omega_0^{5/3} (1+t^2)^{5/6} \over \lambda^2} ,
\label{form1}
\end{equation}
where $t=z/z_R$ and $L_T$ is given by Eq.~(\ref{intlt}). It turns out that the $L_T$ term in Eq.~(\ref{ensf3}) exactly cancels all the $L_T$ terms that appear inside the $L_{mnpq}$ term in Eq.~(\ref{ensf3}) as a result of Eq.~(\ref{form1}).

The integrals of $f_m(K)$ with $m>0$ all give finite results independent of $\kappa_0$ in the limit where $\kappa_0 \rightarrow 0$, and have the form,
\begin{equation}
\int_{-\infty}^{\infty} f_m(K) K\ {\rm d}K \approx G_m {C_n^2 \omega_0^{5/3} (1+t^2)^{5/6} \over \lambda^2} ,
\label{form2}
\end{equation}
where $G_m$ is a numerical constant that only depends on $m$.

\section{Qubit example}
\label{qubit}

\subsection{Solving the IPE}
\label{solveq}

Here we work through a simple example where only modes of the lowest radial index ($r=0$) and with azimuthal indices of the same magnitude $l=\pm q$ are considered. We consider three cases where $q=1,2,3$, respectively. The small number of modes (only two per case) imply a severe truncation. The trace of the truncated density matrix is not in general equal to 1. The truncated density matrix can be normalized to calculate the concurrence \cite{wootters1,wootters2} as a measure of the entanglement. On the other hand, the reduced trace gives an indication of the loss of information to the higher order modes.

For $r=0$ one sets $w_1=w_2=0$ in all the generating functions. Then all the non-dissipative terms (those that contain $S$) in Eq.~(\ref{ensf3}) are equal and cancel each other, so that only the dissipative terms remain.

After evaluating the integrals for $L_{m,n,u,v}(z)$ one finds that, in the limit of large outer scale, the only nonzero elements are
\begin{eqnarray}
& & L_{q,q,q,q}(z) = L_{\overline{q},q,\overline{q},q}(z) \nonumber \\ & = & L_{q,\overline{q},q,\overline{q}}(z) = L_{\overline{q},\overline{q},\overline{q},\overline{q}}(z) =  L_T - A_q h(z)
\label{haa1}
\end{eqnarray}
and
\begin{equation}
L_{q,q,\overline{q},\overline{q}}(z) = L_{\overline{q},\overline{q},q,q}(z) = B_q h(z) ,
\label{haa2}
\end{equation}
where $\overline{q}=-q$; the quantities $A_q$ and $B_q$ are positive constants that only depend on $q$ ($A_1 = 0.03976, B_1 = 0.0007675, A_2 = 0.05588, B_2 = 0.0001213, A_3 = 0.07110, B_3 = 0.00004444$); and $h(z)$ is the same function for all the elements and contains all the dimension parameters,
\begin{equation}
h(z) = \frac{1}{z_R} \left(C_n^2 \omega_0^{2/3} \right) \left(\frac{\lambda}{\pi\omega_0}\right)^{-3} \left(1+\frac{z^2}{z_R^2}\right)^{5/6} ,
\label{haaz}
\end{equation}
where $z_R$ is the Rayleigh range ($\pi\omega_0^2/\lambda$), $\omega_0$ is the radius of the beam waist and $\lambda$ is the wavelength.

The elements in Eq.~(\ref{haa1}), which contain the diverging $L_T$ from Eq.~(\ref{intlt}), are the diagonal elements of $L_{m,n,u,v}(z)$ when treated as a $4\times 4$ matrix. One can therefore view $L_T$ in Eq.~(\ref{haa1}) as being multiplied by an identity matrix. The $L_T$ term in Eq.~(\ref{ensf3}) also represents an identity matrix, but with the opposite sign. This implies that the $L_T$ term in Eq.~(\ref{ensf3}) exactly cancel the $L_T$ elements in Eq.~(\ref{haa1}) for both $L_{m,n,u,v}(z)$ terms in Eq.~(\ref{ensf3}), leaving the final expression without $L_T$. As a result the outer scale drops out of the final expression.

Assuming that the initial state of the density matrix is the singlet Bell state in the OAM basis $(|q\rangle |\overline{q}\rangle - |\overline{q}\rangle |q\rangle)/\sqrt{2}$, one obtains the following solution of the truncated density matrix
\begin{equation}
\rho_{m,n,p,q} = \frac{T}{4} \left[ \begin{array}{cccc}
1-R^2 & 0 & 0 & 0 \\
0 & 1+R^2 & -2R & 0 \\
0 & -2R & 1+R^2 & 0 \\
0 & 0 & 0 & 1-R^2 \\
\end{array} \right] ,
\label{densmat}
\end{equation}
where $mp$ ($nq$) denote the row (column) indices, and where
\begin{eqnarray}
T & = & \exp \left[-2(A_q-B_q) \int_0^z h(z')\ {\rm d}z' \right] \label{tee} \\
R & = & \exp \left[-2B_q \int_0^z h(z')\ {\rm d}z' \right] . \label{er}
\end{eqnarray}
The eigenvalues of the truncated density matrix in Eq.~(\ref{densmat}) are $T(1+R)^2/4$, $T(1-R)^2/4$, $T(1-R^2)/4$ and $T(1-R^2)/4$, which are all positive. The trace of the truncated density matrix is given by $T$, which is a decaying function, since $A_q>B_q$. The trace indicates how much of the information is lost due to coupling to higher order modes that are not represented in the density matrix.

The amount of entanglement for a bipartite qubit system is quantified by the concurrence \cite{wootters1,wootters2}. Normalizing the density matrix by setting $T=1$, one can show that the concurrence for this case is given by
\begin{equation}
{\cal C} = \frac{1}{2}(2R+R^2-1) .
\label{cof}
\end{equation}

\subsection{Fast decay limit}
\label{fdl}

\begin{figure}[ht]
\centerline{\scalebox{1}{\includegraphics{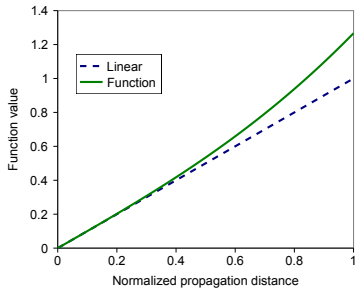}}}
\caption{Comparison of the $t$ and the integral of $h(z)$.}
\label{lyn}
\end{figure}

One can evaluate the integral of $h(z)$ in Eq.~(\ref{haaz}), using
\begin{eqnarray}
\int_0^t \left(1+t_0^2\right)^{5/6}\ {\rm d} t_0 & = & {5 i^{7\over 6} 2^{1\over 6} (i\sqrt{3}-3) \over 12 \pi^{3\over 2}} \Gamma \left({2\over 3}\right)^2  \nonumber \\ & & \times \left(1+t^2\right)^{11\over 12} Q_{-11/6}^{-11/6} \left(it\right) \nonumber \\ & & - {15\sqrt{3} 2^{1\over 3}\over 64\pi} \Gamma \left({2\over 3}\right)^3 \nonumber \\ & = & t + {\rm O} (t^3)
\label{lelik}
\end{eqnarray}
where $t=z/z_R$ and $Q_n^m$ is an associated Legendre function of the second kind \cite{as2}. For small $t$ the result of the integral in Eq.~(\ref{lelik}) is approximately equal to $t$. In Fig.~\ref{lyn} the result in Eq.~(\ref{lelik}) is compared with the line given by $t$. One can see that for $t \lesssim 1/3$ the result of Eq.~(\ref{lelik}) could be fairly well approximated by $t$. What this implies is that for propagation distances much shorter than the Rayleigh range the integral of $h(z)$ can be approximated by,
\begin{equation}
\int_0^z h(z')\ {\rm d}z' \approx 0.5928 \left( \frac{\omega_0}{r_0} \right)^{5/3} ,
\label{hprox}
\end{equation}
where $r_0 = 0.185(\lambda^2/C_n^2/z)^{3/5}$, which is the Fried parameter. Thus all the dimension parameters are combined into $\omega_0/r_0$. If the entanglement completely decays over distances much shorter than the Rayleigh range --- a situation, which we call the fast decay limit --- one can use Eq.~(\ref{hprox}) and thus express the entire behavior simply as a function of $\omega_0/r_0$. For larger values of $t$ the behavior becomes more complicated as shown in Eq.~(\ref{lelik}). In that case the behavior does not only depend on $\omega_0/r_0$ but also on the equivalent of $\omega_0/r_0$ with $z$ replaced by $z_R$.

\subsection{Comparison}
\label{vglk}

\begin{figure}[ht]
\centerline{\scalebox{1}{\includegraphics{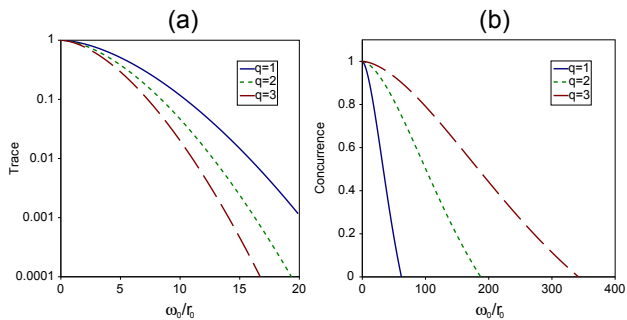}}}
\caption{Plots of (a) the trace of the truncated density matrix and (b) the concurrence for a biphoton, initially in the singlet Bell-state, in terms of two OAM states with $l=\pm q$, for $q=1,2,3$, as a function of $\omega_0/r_0$.}
\label{conc}
\end{figure}

Previous analyses of the decoherence of OAM entanglement due to atmospheric scintillation \cite{qturb1,qturb3,qturb4} obtained results that only depend on $\omega_0/r_0$. So to compare our results with their results we need to consider our results in the fast decay limit. This is done by substituting Eq.~(\ref{hprox}) into $R$ and $T$ in Eqs.~(\ref{tee}) and (\ref{er}), and then substitute $R$ into Eq.~(\ref{cof}). We plot the resulting trace $T$ and the resulting concurrence ${\cal C}$ as functions of $\omega_0/r_0$ for $q=1,2,3$ in Figs.~\ref{conc}(a) and \ref{conc}(b), respectively.

From the curves for the trace in Fig.~\ref{conc}(a) one can see that modes with higher OAM are scattered more rapidly into other modes than those with lower OAM. On the other hand, from Fig.~\ref{conc}(b) we see that modes with higher OAM retain their entanglement for longer distances than those with lower OAM. These conclusions agree qualitatively with previous work \cite{qturb1}, however, the scattering into other modes occurs at a scale where $\omega_0/r_0 \gg 1$ and the entanglement lasts for at least two orders of magnitude longer than was found in Ref.~\cite{qturb1}. Here the slowness of the decay in the concurrence is a result of the smallness of the values of the $B_q$'s. (While the $A_q$'s come from the autocorrelation functions of the OAM modes, the $B_q$'s come from the cross-correlations of modes with different azimuthal indices, which give much smaller overlaps with the power spectral density.) From these results it appears that the effect of scattering and the implied loss of photons in the desired OAM modes may turn out to be a more significant challenge for free-space quantum communication than the decoherence of OAM entanglement. On the other hand, the effects of the severe trunction may imply that these results are but a poor reflection of what really would happen in this scenario.

\section{Discussion}
\label{disc}

The loss of entanglement due to decoherence is a challenge that confronts the development of quantum communication systems. It is necessary to be able to predict the propagation scale over which this decoherence takes place before a successful free-space quantum communication can be designed. Previous attempts to make such predictions \cite{qturb1} were based on certain assumptions \cite{turboam1}, the effects of which were perhaps not completely clear. Intuitively, the assumption that one can represent the atmosphere by a single phase screen sounds reasonable, and so does the assumption that one can represent the strength of the turbulence by the Fried parameter. However, carefully considering the effect of turbulence, one realizes that some physical effects of the scintillation process is lost as a result of these assumptions.

Turbulence is a cascaded process. The random index fluctuations cause a phase modulation of the traversing optical beam. Directly after an initial random phase modulation, the amplitude of the optical beam is unaffected. However, during subsequent propagation the phase distortion is partially transferred into an amplitude distortion. This mixture of phase and amplitude distortion receives further phase modulations as the beam passes through the random medium. The random phase modulation and the propagation both play important roles in the process that produces the scintillated beam. Without the propagation part of the process the phase modulation will never be converted into amplitude scintillation. Since OAM modes have particular phase and amplitude characteristics, both the phase and the amplitude are important to give the correct scattering coefficients.

The formulation that is presented in this paper takes into account both the random phase modulation and the propagation and therefore gives the required effect on the phase and the amplitude of the modes.

The effect of having a single phase screen where the turbulence is completely characterized by the Fried parameter is shown by the fast decay limit, discussed in Section~\ref{fdl}. Only in this limit can the effect of the turbulence be completely described by the Fried parameter. Beyond this limit the scale parameters in the problem starts to contribute in a way that cannot be combined into the Fried parameter. In other words, different combinations of these scale parameters can give different predictions even when the Fried parameter remains fixed. The implication is that, when one models the quantum scintillation process by a single phase screen in terms of the Fried parameter one tacitly assumes the fast decay limit and it is debatable whether the fast decay limit is valid or even useful for a free-space quantum communication system.

\section{Summary}
\label{summ}

We use the transformation of the momentum space wave functions of OAM modes after an infinitesimal propagation through a turbulent atmosphere to derive an IPE for the decoherence of OAM entangled biphoton states. The resulting set of first order differential equations is used to study the evolution of a severely truncated density matrix where the initial state of a biphoton is an entangled qubit in the OAM basis. The results are compared with previous results in the literature.

\appendix

\section{Generating function for OAM modes}
\label{oamgen}

The Laguerre-Gaussian (LG) modes, which are solutions of the paraxial wave equation, are given in terms of normalized coordinates by
\begin{eqnarray}
M^{LG}_{r,l}(u,v,t) & = & {\cal N} {(u \pm iv)^{|l|} (1+it)^r \over (1-it)^{r+|l|+1}} \exp \left( {u^2+v^2 \over it-1} \right) \nonumber \\
& & \times L_r^{|l|} \left( {2 (u^2+v^2)\over 1+t^2} \right) ,
\label{lgm}
\end{eqnarray}
where the normalized coordinates are given by $u=x/\omega_0$, $v=y/\omega_0$ and $t=z/z_R=z\lambda/\pi\omega_0^2$, in terms of the initial radius of the mode profile $\omega_0$ and the Rayleigh range $z_R$; $r$ is the radial index (a non-negative integer); $l$ is the azimuthal index (a signed integer); the $\pm$ sign is given by the sign of $l$; ${\cal N}$ is a normalization constant given by
\begin{equation}
{\cal N} = \left[ {r!2^{|l|+1} \over \pi (r+|l|)!} \right]^{1/2}
\label{lgn}
\end{equation}
and $L_r^{|l|}$ represents the associate Laguerre polynomials, which can be obtained from the generating function,
\begin{equation}
g^{|l|}(x,w) = {1\over (1-w)^{1+|l|}} \exp\left({-xw\over 1-w} \right) ,
\label{lgen}
\end{equation}
by computing its $r$-th derivative,
\begin{equation}
L_r^{|l|}(x) = {1\over r!} \left. {{\rm d}^r g^{|l|}(x,w) \over {\rm d} w^r} \right|_{w=0} .
\label{lgfunk}
\end{equation}

The azimuthal index $l$ of the LG modes represent the amount of OAM that each photon in such an optical modes carries. Therefore, the LG modes act as an OAM basis in quantum optics. The LG mode functions are the position space wave functions of the OAM states and their Fourier transforms are the corresponding momentum space wave functions.

In the paraxial limit the LG modes are treated as two-dimensional functions of the transverse coordinates $u$ and $v$, and these two-dimensional functions change as a function of the normalized propagation distance $t$.

One can use a generating function for the OAM basis functions (LG modes) to evaluate the integrals that contain OAM momentum space wave functions. Such integrals only have to be solved once and afterward the solutions for particular OAM modes are generated using derivatives. Since it is always easier to compute derivatives than to solve integrals this represent a reduction in the computation effort that needs to be performed.

The generating function for the LG modes in normalized coordinates is given by
\begin{eqnarray}
G & = & \sum_{n,m=0}^{\infty} \frac{1}{m!} L_n^m\left({2(u^2+v^2)\over 1+t^2}\right) \left[{w(1+it)\over 1-it}\right]^n \nonumber \\
& & \times {\left[(u+iv)p + (u-iv)q\right]^m \over (1-it)^{1+m}} \nonumber \\
& = & {1 \over \Omega(t,w)} \exp \left[ {(u+iv)p \over \Omega(t,w)} + {(u-iv)q \over \Omega(t,w)} \right. \nonumber \\
& & - \left. {(1+w) (u^2+v^2) \over \Omega(t,w)} \right] ,
\end{eqnarray}
where $\Omega(t,w) = 1-w-it-iwt$. The parameters $p$, $q$ and $w$ are used to generate particular LG modes in the following way,
\begin{equation}
M^{LG}_{r,l}(u,v,t) = \left\{ \begin{array}{lcc}
{\cal N} \left[ \frac{1}{r!} \partial_w^r \partial_p^{|l|} G \right]_{w,p,q=0} & {\rm for} & l>0 \\
{\cal N} \left[ \frac{1}{r!} \partial_w^r  G \right]_{w,p,q=0} & {\rm for} & l=0 \\
{\cal N} \left[ \frac{1}{r!} \partial_w^r \partial_q^{|l|} G \right]_{w,p,q=0} & {\rm for} & l<0 , \\
\end{array} \right.
\label{ftgen}
\end{equation}
where $r$ and $l$ represent the radial and azimuthal indices, respectively, and ${\cal N}$ is the normalization constant given in Eq.~(\ref{lgn}).

The integrals in Eqs.~(\ref{defs}) and (\ref{defw}) contain the Fourier transform of the LG modes. For this purpose we need the Fourier transform of the generating function, which is given by\begin{eqnarray}
{\cal F} \{G\} & = & \frac{\pi}{1+w}\exp \left[ { i\pi (a+ib)p \over 1+w } + {i\pi (a-ib)q \over 1+w } \right. \nonumber \\ & & \left. - {\pi^2 (a^2+b^2)\Omega(t,w) \over 1+w } \right]
\label{msgf}
\end{eqnarray}
where $a$ and $b$ are normalized spatial frequency components that are related to $k_x$ and $k_y$ through
\begin{equation}
k_x = {2\pi a\over \omega_0} ~~~~~~~~~ k_y = {2\pi b\over \omega_0} .
\end{equation}
The Fourier transforms of particular LG modes are obtained using the same procedure given in Eq.~(\ref{ftgen}).

\section{Ensemble average} 
\label{ensemb}

As mentioned in Sec.~\ref{eom}, the refractive index fluctuations produced by a turbulent atmosphere are small compared to the average refractive index of air, $\delta n = n-1 \ll 1$, which leads to the fact that one can separate the propagation through a turbulent atmosphere into two parts: free-space propagation and the random phase modulation. The random phase function for the latter step is obtained by integrating the refractive index fluctuations through a thin sheet of atmosphere along the propagation direction,
\begin{eqnarray}
\theta(x,y) & = & k \int_{z_0-\delta z/2}^{z_0+\delta z/2} \delta n(x,y,z) {\rm d} z \nonumber \\
& \approx & k\ \delta z\ \delta n(x,y,z_0) ,
\end{eqnarray}
where, in the last line we took the limit $\delta z \rightarrow 0$. Replacing the refractive index fluctuation with its Fourier expansion, we obtain
\begin{eqnarray}
\theta(x,y,z_0) & = & k \delta z \int \exp[-i(k_x x + k_y y + k_z z_0)] \nonumber \\
& & \times N_n({\bf k})\ {{\rm d}^3 k\over (2\pi)^3} ,
\end{eqnarray}
where $N_n({\bf k})$ is the three-dimensional spatial spectrum of index fluctuations. One can now define a two-dimensional spectrum for the accumulated index fluctuations over a thin sheet of atmosphere as follows,
\begin{equation}
N_n({\bf K},z) = \int \exp(-i k_z z) N_n({\bf k})\ {{\rm d} k_z\over 2\pi} ,
\label{spek2d}
\end{equation}
where $N_n({\bf K},z)$ is the two-dimensional spectrum, which depends on the $z$ position of the thin sheet. The three-dimensional spectrum of the refractive index fluctuations can be expressed in terms of its three-dimensional power spectral density, which follows from the correlation function of the index fluctuations and which represents the model for the turbulence,
\begin{equation}
N_n({\bf k}) = \tilde{\chi}_n({\bf k}) \left[ {\Phi_0({\bf k})\over\Delta_k^3} \right]^{1/2} ,
\label{nspek}
\end{equation}
where $\tilde{\chi}({\bf k})$ is a normally distributed random complex spectral function and $\Delta_k$ is its spatial coherence length in frequency domain. The latter is inversely proportional to the outer scale of the turbulence. Since the refractive index fluctuation $\delta n$ is an asymmetric real-valued function, we have that $\tilde{\chi}^*({\bf k})=\tilde{\chi}(-{\bf k})$. Furthermore, the autocorrelation function of the random function is given by
\begin{equation}
\langle \tilde{\chi}({\bf k}_1) \tilde{\chi}^*({\bf k}_2) \rangle = \left( 2\pi\Delta_k \right)^3 \delta_3 ({\bf k}_1-{\bf k}_2) .
\label{verwrand}
\end{equation}

In Eq.~(\ref{ensf}) we find ensemble averages inside double $z$-integrals. Substituting Eqs.~(\ref{spek2d}) and (\ref{nspek}) into these ensemble averages, using Eq.~(\ref{verwrand}) to evaluate the ensemble average, one obtains
\begin{eqnarray}
& & \int_{z_0}^{z} \int_{z_0}^{z_1} \langle N_n({\bf K}_1,z_2) N_n^*({\bf K}_2,z_1) \rangle\ {\rm d} z_2\ {\rm d} z_1 \nonumber \\
& = & (2\pi)^2 \delta({\bf K}_1-{\bf K}_2) \int \int_{z_0}^{z} \int_{z_0}^{z_1} \Phi_0({\bf k}_1) \nonumber \\
& & \times \exp \left[i k_z (z_1-z_2) \right]\ {\rm d} z_2\ {\rm d} z_1\ {{\rm d} k_z\over 2\pi}
\label{verw1}
\end{eqnarray}
Setting $z=z_0+dz$, we evaluate the two $z$-integrals
\begin{eqnarray}
& & \int_{z_0}^{z_0+dz} \int_{z_0}^{z_1} \exp \left[i k_z (z_1-z_2) \right]\ {\rm d} z_2\ {\rm d} z_1 \nonumber \\
& = & {1-\cos(k_z dz)\over k_z^2} + i {\sin(k_z dz)-k_z dz\over  k_z^2} .
\label{zint}
\end{eqnarray}
The power spectral density $\Phi_0({\bf k}_1)$ is always even in $k_z$. Therefore, the imaginary part of Eq.~(\ref{zint}), being odd in $k_z$, does not contribute to the final expression. So we have
\begin{eqnarray}
& & \int_{z_0}^{z} \int_{z_0}^{z_1} \langle N_n({\bf K}_1,z_2) N_n^*({\bf K}_2,z_1) \rangle\ {\rm d} z_2\ {\rm d} z_1 \nonumber \\
& = & (2\pi)^2 \delta({\bf K}_1-{\bf K}_2) \int \Phi_0({\bf k}_1) \nonumber \\ & & \times \left[ {1-\cos(k_z dz)\over k_z^2} \right] \ {{\rm d} k_z\over 2\pi} .
\label{verw2}
\end{eqnarray}
Due to the fact that the refractive index variations are very small, the light that propagates through the turbulent atmosphere remains unchanged over distances much longer than the correlation distance of the turbulent medium. One can therefore assume that $dz$ is much larger than this correlation distance. As a result the function inside the square-brackets in Eq.~(\ref{verw2}) acts like a Dirac delta function, so that one can substitute $k_z=0$ in $\Phi_0$ and pull it out of the $k_z$-integral. The integral can then be evaluated to give
\begin{eqnarray}
& & \int_{z_0}^{z_1} \int_{z_0}^{z} \langle N_n({\bf K}_1,z_2) N_n^*({\bf K}_2,z) \rangle\ {\rm d} z_2\ {\rm d} z \nonumber \\
& = & 2\pi^2 dz \delta({\bf K}_1-{\bf K}_2) \Phi_1({\bf K}_1) ,
\end{eqnarray}
where we defined $\Phi_1({\bf K}_1)=\Phi_0({\bf K}_1,0)$.


\end{document}